\documentstyle[12pt,aps,epsf]{revtex}
\begin{document}
\thispagestyle{empty}
\title{
\hspace{4.in} {\normalsize DOE/ER/40427-08-N96} \\
\hspace{4.in} {\normalsize DOE/ER/40561-263-INT96-19-04}\\
\hspace{4.in} {\normalsize ADP-96-18/T221} \\
\vspace{0.4cm}
Electromagnetic Gauge Invariance of the Cloudy Bag Model}

\author{Gerald A. Miller$^a$\footnote{e-mail:
miller@phys.washington.edu}, and Anthony W. Thomas$^b$\footnote{e-mail:
athomas@physics.adelaide.edu.au}}
\address{
national Institute of Nuclear Theory, Box 351550, 
University of Washington, Seattle, WA 98195-1560, USA}
\address{$^a$ Permanent address,
Department of Physics, Box 351560, University of Washington,
Seattle, Washington 98195-1560,  USA}

\address{$^b$ Permanent address,
Department of Physics and Mathematical Physics,\\
and Special Research Centre for the Subatomic Structure of Matter, \\
University of Adelaide, Adelaide, South Australia 5005 AUSTRALIA}

\maketitle
\begin{abstract}

We examine the question of the gauge invariance of electromagnetic form factors
calculated within the cloudy bag model. 
One of the assumptions of 
the model is that electromagnetic form factors are most 
accurately evaluated in the Breit frame. This feature is used to show that
gauge invariance is respected in this frame.
\end{abstract}

\newpage

\section{Introduction}
The present experimental efforts \cite{expt} to determine the strangeness 
content of the nucleon have led to considerable interest in 
nucleon models which include pionic
and kaonic clouds. For example, one way to
model the strangeness content of the nucleon is to regard it as having a
$\bar K \Lambda$ component\cite{ST}.

In the cloudy bag model \cite{CBM,CBM2}, the pion field required by chiral
symmetry is quantized and
coupled to the quarks in an MIT bag
\cite{MIT}. This leads to a model that
includes relativistic quark wave functions and quark confinement 
as well as respecting chiral symmetry.
The cloudy bag model was applied to computing
$\pi$- \cite{CBM,PIN} and  K-nucleon scattering \cite{KN}; 
baryon electromagnetic \cite{EM,HYP} and
axial form factors \cite{AX}; M1 radiative decays of mesons of both
light and heavy flavor \cite{MES} and most recently deep-inelastic
scattering \cite{HOL}. Generally good agreement with experiment
was obtained. Moreover, the model provided a framework for 
nuclear physics which included the old meson cloud physics along
with the new quark degrees of freedom in a consistent
theoretical framework \cite{CBM2}.

Recently, Musolf and Burkardt \cite{MB} and Koepf and Henley 
\cite{KH} have argued that the standard computation of the nucleon
electromagnetic form factors within such composite models 
is not gauge invariant.
An attempt was made to remedy this perceived problem
by adding a contact interaction, as suggested by
Gross and Riska \cite{GR}, or by applying the minimal substitution
prescription of Ohta \cite{Ohta}.  These ``extra" terms can be 
relatively large, especially when applied to the kaon cloud.

Of course, questions of local gauge invariance arise quite generally in
nuclear physics, where one often works with effective meson-nucleon
interactions involving momentum dependent vertex functions. We recall that the
cloudy bag model electromagnetic form factors are most accurately calculated in
the Breit frame in which the initial momentum of the nucleon is $- \vec q/ 2$,
the final momentum is $\vec q /2$ and the energy transferred to the target
nucleon vanishes. The use of this special frame was necessary because the
evaluations of the model were not covariant and because recoil effects were not
included. The problems associated with this deficiency can be reduced by
limiting the energy of each 
nucleon to $\sqrt{\vec{q}^2/4 + m^2}$. Once obtained,
the form factors can be used in any frame, under the restriction
that $-q^2/4m^2$ be small.

Although the underlying, quark-level Lagrangian respects electromagnetic
gauge invariance, the CBM does not respect gauge invariance in all frames. In
this paper we show that the CBM respects gauge invariance in the Breit frame
and, therefore, that in this frame no extra terms need to be added for
calculating the nucleon electromagnetic form factors. In order to prove this
result one must include the vertex correction illustrated in Fig. 1(b), which
was omitted in Ref. [14].

The use of the Breit frame was essential to the CBM (as to all the older,
static source, meson theories [17, 18]]) where the source has known internal
structure - the MIT bag [5]. In a static model in contrast, there is no
difference between a nucleon of momentum $\vec 0$ and $- \vec q/2$, so that the
choice of the Breit frame was implicit in the identification of $G_E(Q^2)$ with
the matrix element of $j^0$, while the matrix element of $\vec j$ 
was identified with $G_M(Q^2)$.

\section{Calculations}

Calculations of electromagnetic form factors, $\Gamma^\mu$, and nucleon
self-energies, $\Sigma$, obey electromagnetic gauge invariance if 
the Ward identity for nucleons of mass m:
\begin{equation}
q_\mu \Gamma^\mu (p',p) = S^{-1} (p') - S^{-1} (p) \,\, ,
\end{equation}
is satisfied.  
Here the inverse propagator of the nucleon is
$S^{-1} (p) = \gamma\cdot p  - m - \Sigma (p)$ and 
$p' = p + q$. As we have already explained,
the CBM is not covariant.
All calculations of electromagnetic form factors were made in
time-ordered perturbation theory with on-mass-shell nucleons. 
Indeed there was no discussion on how to continue the model 
off-mass-shell. Thus, the
only legitimate concern over gauge invariance within the model, 
is that Eq. (1) is satisfied
when evaluated between on-mass-shell spinors $\bar{u}(p')$ and u(p).
That is, we need to check whether 
\begin{equation}
q_\mu \bar{u}(p') \Gamma^\mu (p',p) u(p) = \bar{u}(p') \left[ 
S^{-1} (p') - S^{-1} (p) \right] u(p).
\end{equation}

It is useful to start by evaluating the cloudy bag model (CBM) value of 
$\Sigma_{CBM} (p') - \Sigma_{CBM}(p)$.
We use the loop integral formulation of the cloudy bag model, introduced by 
Koepf and Henley, to evaluate these quantities. Then 
\begin{equation}
\Sigma (p) = 3 ig^2 \int {d^4k \over (2 \pi)^4} \gamma_5 f(k) S 
(p-k) f(k) \gamma_5 \Delta (k) ,
\end{equation}
where $\Delta$ is the pion propagator, 
$f(k)$ is the cloudy bag model form factor ($f(k)=3 j_1(|\vec k|R)
/|\vec k|R$, with R the radius of the bag)
and $g$ is the pion nucleon coupling 
constant. We use the pseudoscalar form of the cloudy bag model, but the
pseudovector version yields the same result,
as we shall see.

\begin{figure}[t]
\noindent
\epsfysize=0.9in
\hspace{1.0in}
\epsffile{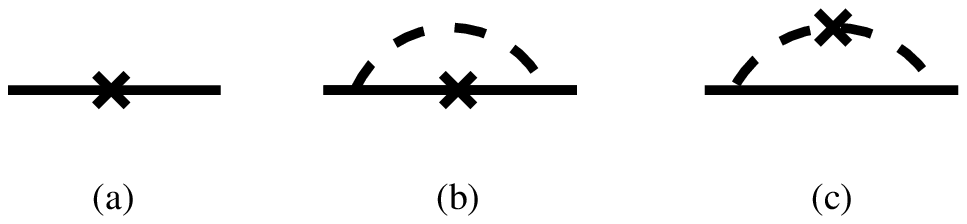}
\begin{center} Figure 1: Contributions to the nucleon electromagnetic
form factors.
\end{center}
\end{figure}

We have already noted that the CBM involves only
on-mass-shell nucleons. The effects of $N \bar{N}$ 
pairs were not included in the model -- they were supposed to be
suppressed by form factors for the composite particle. 
Thus, for the bare nucleon propagator it is consistent with the CBM to
use the form
\begin{equation}
S_0(p) \rightarrow {1\over 2E(p)} {\gamma^0 E(p)-\vec\gamma\cdot \vec{p} + m 
\over p^0 - E(p) + i\epsilon} 
= {m \over 2E(p)} {u(p) \bar{u} (p) \over p^0 - E (p) + i\epsilon}
\end{equation}
where 
$E(p) = \sqrt{\vec{p}\,^2 + m^2}$. The pseudovector and pseudoscalar 
interactions have the same matrix elements between
on-shell spinors, so this is the equality mentioned above. From Eq. (4)
we see that all that enters the dressed nucleon propagator in the CBM is
the matrix element of $\Sigma$ between on-shell spinors. Therefore the
self-energy in the CBM was effectively the Dirac scalar quantity 
$\Sigma_{CBM}(p)$: 
\begin{equation}
\Sigma_{CBM}(p) = 3ig^2 \int {d^4k \over (2 \pi)^4} {f^2 (k) \Delta (k) 
\over 2E(\vec{p} - \vec{k})} {\bar{u}(p) \gamma\cdot k u(p) \over p^0 - k^0 -E(\vec{p}- 
\vec {k}) + i\epsilon}. 
\end{equation}
We do the $dk^0$ integral over the lower half plane, so that $k^0 \rightarrow 
\omega_k = \sqrt{\vec{k} ^2 + m_\pi^2}$ .
A simple evaluation shows that
\begin{equation}
\Sigma_{CBM}(p) = \Sigma_{CBM}(E(\mid \vec{p}\mid),\vec{p}) = 
{E(\mid\vec{p}\mid) \over m} I_1 
(\mid\vec{p} \mid) + {I_2 ((\mid \vec{p} \mid )\over m} 
\end{equation}
where
\begin{equation}
I_1(\mid \vec{p} \mid) = {-3g^2 \over 4} \int {d^3k \over (2 \pi)^3} 
{f^2 (\mid \vec{k} \mid ) \over E(\mid \vec{p} - \vec{k} \mid)} 
{1 \over \omega _k + E (\mid\vec{p} - \vec{k}\mid)
 - E (\mid \vec{p} \mid )} \;\; .
\end{equation}
and
\begin{equation}
I_2(\mid \vec{p} \mid) = {-3g^2 \over 4} \int {d^3k \over (2 \pi)^3} 
{f^2 (\mid \vec{k} \mid ) \over E(\mid \vec{p} - \vec{k} \mid)} 
{\vec k\cdot \vec p
\over \omega _k + E (\mid\vec{p} - \vec{k}\mid) - E (\mid \vec{p} \mid )} \;\; .
\end{equation}

In the Breit frame $\mid \vec{p}\;' \mid = \mid \vec{p} \mid = \mid \vec{q} 
\mid/ 2$ and $E(\vec{p}\; ' ) = E (\mid \vec{p} \mid ) = 
\sqrt{{\vec{q}^2 \over 4} + m^2}$. Since Eq. (6) shows that 
the self energy depends only on
$\mid \vec{p}\mid$, it is then clear that 
\begin{equation}
\Sigma_{CBM} (p') - \Sigma_{CBM} (p) = 0 \;\; ,
\label{good}
\end{equation}
in this frame.

Considering first the bare vertex, $\gamma^{\mu}$, shown in Fig. 1(a),
we see that its contraction with $q_{\mu}$ cancels the term $\gamma
\cdot p' - \gamma \cdot p$ on the right hand side of Eq. (1).
This means that all that is needed to demonstrate the electromagnetic gauge 
invariance of the CBM in the Breit frame is to show that $q_\mu \bar{u}(p')
[\Gamma_{CBM}^\mu (p', p) 
- \gamma^\mu ] u(p) = 0$.

There are two contributions to $\Gamma_{CBM}^\mu (p',p) - \gamma^\mu$. 
There is a vertex 
contribution $V^\mu (p',p)$ in which the photon hits the nucleon while the 
pion is in the air (Fig. 1(b)), and the direct term $\Gamma_\pi ^\mu (p',p)$ 
in which the photon couples to a charged pion (Fig. 1(c)). 
Then $\bar{u}(p')[ \Gamma^\mu - \gamma^\mu ]u(p) = V^\mu + \Gamma_\pi^\mu$.

It is straightforward to obtain
\begin{equation}
q_\mu V^\mu (p',p) = ig^2\bar{u}(p')\int {d^4k \over (2 \pi)^4} \gamma_5
f(k) S(p-k) \gamma\cdot q S(p'-k) f(k)\gamma_5 u(p) \Delta(k^2) .
\end{equation}
But
\begin{equation}
S(p-k) \gamma\cdot q S(p'-k) = S(p-k) - S(p' -k) \;\; ,
\end{equation}
so that
\begin{equation}
q_\mu V^\mu = {1 \over 3} (\Sigma (p) - \Sigma(p')) \;\; ,
\end{equation}
and from Eq.(9) we have
\begin{equation}
q_\mu V^\mu = 0 \;\; .
\end{equation}

Now consider $q_\mu\Gamma_\pi^\mu$. The Feynman graph is given by
\begin{eqnarray}
q_\mu \Gamma_\pi^\mu (p',p) = i2g^2 \int {d^4k \over (2 \pi)^4} 
\bar{u}(p') \gamma_5 f(k-q / 2) S(p-k+q/2)f(k+q/2) \gamma_5 u(p) \\
\Delta (k-q/2) \Delta (k+q/2) ((k+q/2)^2 - (k-q/2)^2),
\end{eqnarray}
in which we have replaced the original integration variable $k$ by $k-q/2$.  
Then use Eq.(4) in the Breit frame ($q^0 = 0$) to obtain
\begin{eqnarray}
q_\mu \Gamma_\pi^\mu (p',p) = {2ig^2 \over 4} \int {d^4k \over (2\pi)^4} 
{\bar{u} (p') \gamma\cdot k u(p) \over E(k) (E(q/2) - k^0 - E(k))} \cr
\Delta (k-q/2) f(k-q/2) \Delta (k+q/2) f(k+q/2) ((k+q/2)^2 - (k-q/2)^2)
.
\end{eqnarray}
Consider the $d^3k$ integral for fixed $k^0$, with $\gamma\cdot k
 = (\gamma^0 k^0 -
\vec{\gamma}\cdot \vec{k})$. 
The coefficient of the $\gamma^0 k^0$ term vanishes 
since it is multiplied by an odd function of $\vec{k}$. Furthermore, the
integral $d^3k$ of 
the $\vec{\gamma} \cdot \vec{k}$ term must be proportional 
to $\vec{\gamma} \cdot \vec{q}$ 
because $\vec p = \vec q/2 = -\vec{p}\,'$. However, 
$\bar{u}(\vec q /2) \vec \gamma \cdot
\vec {q} u(- \vec{q} /2) = 0$. Thus we have $q_\mu \Gamma_\pi^\mu 
(p',p) = 0$.  
Other terms, such as the anomalous magnetic coupling to the 
intermediate nucleon, the influence of intermediate $\Delta$ states
and the effects of the pion-quark-photon contact interaction \cite{87}
are individually
gauge invariant. This means that the 
proof that the CBM respects 
electromagnetic gauge invariance in the Breit frame is complete.

\section{Discussion}


We recognize that the utility of the result proved above is limited. 
It would be 
better to use a fully covariant model so that one could verify the 
Ward identity in any frame. No such model involving quarks, mesons and 
photons exists. The present work is limited to the
goal of showing that the cloudy bag model was not grossly incorrect.

\newpage
\centerline{ACKNOWLEDGEMENTS}

We thank  T.\ E.\ O.\ Ericson, E.\ Henley, G.\ Krein, D.\ Lu, and 
A.\ G.\ Williams for useful discussions and the 
national Institute for Nuclear Theory for its support. This work was supported 
in part by U.S. Department of Energy grant DE-FG06-88ER40427 and by the
Australian Research Council.

\newpage

\end{document}